\pdfoutput=1

\documentclass[twocolumn,showpacs,amssymb,aps,nofootinbib,floatfix,groupedaddress]{revtex4-1}

\usepackage{epsfig}
\usepackage{hyperref}
\usepackage{comment}
\usepackage{amsmath}
\usepackage{graphicx,color}
\begin{document} 
\hbadness=10000

\title{Azimuthal angle dependence of the charge imbalance from charge 
conservation effects}

\author{Piotr Bo\.zek}
\email{piotr.bozek@fis.agh.edu.pl}
\affiliation{AGH University of Science and Technology, Faculty of Physics and
Applied Computer Science, al. Mickiewicza 30, 30-059 Krakow, Poland}

\begin{abstract}
The experimental search for the chiral magnetic effect in heavy-ion collisions
 is based on charge
 dependent correlations between emitted particles. Recently, a  sensitive
observable comparing  event-by-event distributions of the charge splitting 
projected on the directions along  and perpendicular to the direction
 of the elliptic flow has been proposed. The results of a
 3+1-dimensional hydrodynamic model show that the preliminary experimental
 data of the STAR Collaboration can be explained as due to  background 
effects, such as resonance decays and local charge conservation in the 
particle production. A related observable based on the third order harmonic flow is proposed to further investigate such background effects in charge dependent correlations.
\end{abstract}

\date{\today}

%\pacs{25.75.-q, 25.75Gz, 25.75.Ld}


\maketitle

%%%%%%%%%%%%%%%%%%%%%%%%%%%%%%%

\section{Introduction \label{sec:intro}}

The possibility of  creating 
topological domains with a non-zero topological charge in the dense
 matter created in heavy-ion collisions has been proposed
 \cite{Kharzeev:2004ey,Fukushima:2008xe,Kharzeev:2015znc}. In the presence of
a magnetic field, as in the early stages of the collision,
 this would induce an event-by-event charge separation  between emitted
particles, the chiral magnetic effect. The
charge splitting between same-sign and unlike-sign pairs 
in a specific  
correlator has been proposed as a sensitive observable
 to discover the presence of topological domains 
 \cite{Voloshin:2004vk}.

The 
chiral magnetic effect has been searched for  experimentally in heavy-ion collisions
\cite{Abelev:2009ac,Adamczyk:2013kcb,Abelev:2012pa,Khachatryan:2016got,Adamczyk:2013hsi}. One observes a charge dependence in two-particle correlations
 with respect to 
the second order harmonic flow. On the other hand, many of 
these observations are quantitatively explained as due to standard charge correlations
 present in  particle production 
\cite{Bzdak:2010fd,Wang:2009kd,Schlichting:2010qia,Bzdak:2012ia}. 
The simplest example of such correlation is given by the correlation
 between decay products of resonances. The phenomenon is generic 
and any particle formation mechanism should obey charge conservation 
constraints. Local charge conservation leads to correlation between
 unlike-signed particles in phase-space \cite{Bass:2000az}. As a result
 two-particle 
correlation in rapidity and/or azimuthal angle show a definite charge dependence
\cite{Cheng:2004zy,Bozek:2004dt,Bozek:2012en}.

Recently, a new observable has been proposed as a measure of charge dependent splitting along the
direction of the magnetic field.
The event-by-event distribution of the charge splitting projected on the direction of the magnetic field (perpendicular to the elliptic flow direction)
 is compared between real events and events with randomized charges 
 \cite{Ajitanand:2010rc}. A more sensitive observable is constructed as
 the ratio of the event-by-event distributions of charge splitting 
perpendicular and along the direction of the elliptic 
flow \cite{Magdy:2017yje}. It has been noticed
 that the effect observed in preliminary data of the STAR 
Collaboration \cite{talkRoySTAR} is qualitatively different from models 
without a chiral magnetic effect.

In this paper I calculate the distribution of the charge splitting in  a
3+1-dimensional hydrodynamic model for  A+A and p+Pb collisions.
I show that in simulations involving correlations between charged particles
 from resonance decays or local charge conservation \cite{Bozek:2012en}
the experimental data are qualitatively reproduced, i.e. the 
charge splitting distribution is wider out-of-plane than in-plane. 
A similar  observable is proposed involving the third order event-pane. 
In this new observable any charge dependence of the correlations comes 
solely from the triangular flow, without any 
contribution from the chiral magnetic 
effect. This observable  could 
give as an additional test of possible background 
effects in the search for QCD topological domains in heavy-ion collisions.

\section{Event-by-event distribution of the charge splitting}

\begin{figure}[tb]
%\begin{center}
%\vskip -60mm
\includegraphics[width=.45 \textwidth]{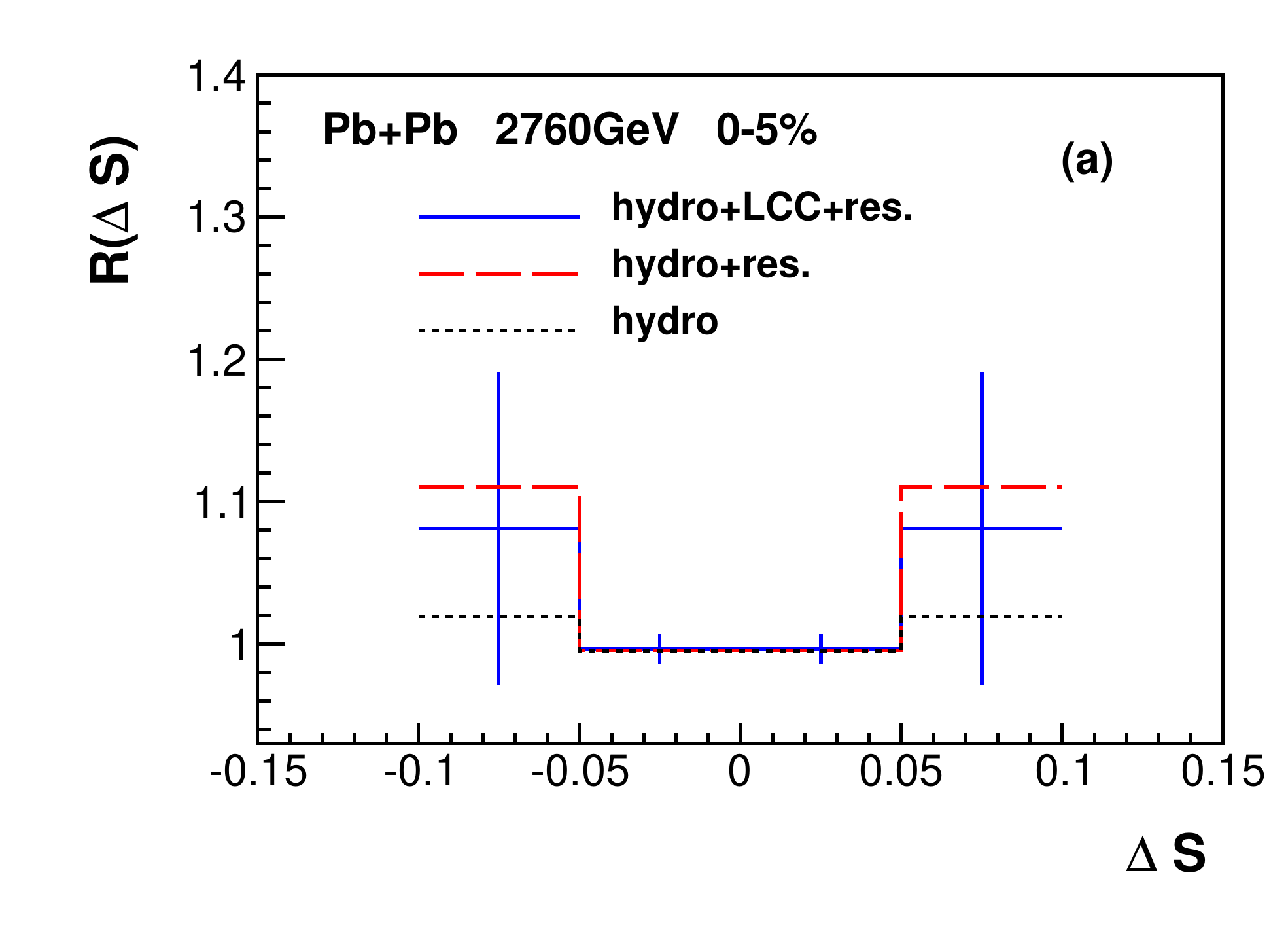}

\includegraphics[width=.45 \textwidth]{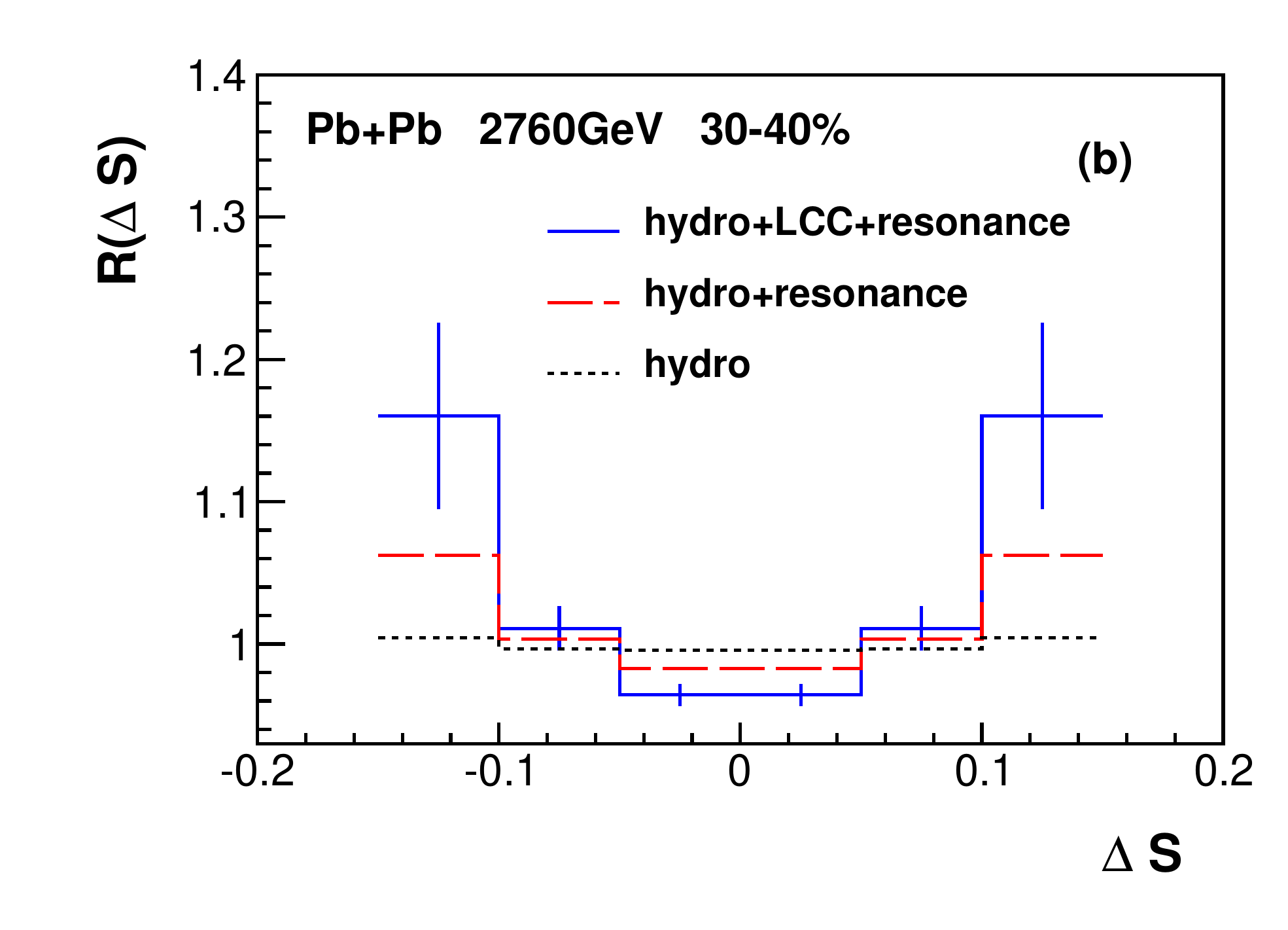}

\includegraphics[width=.45 \textwidth]{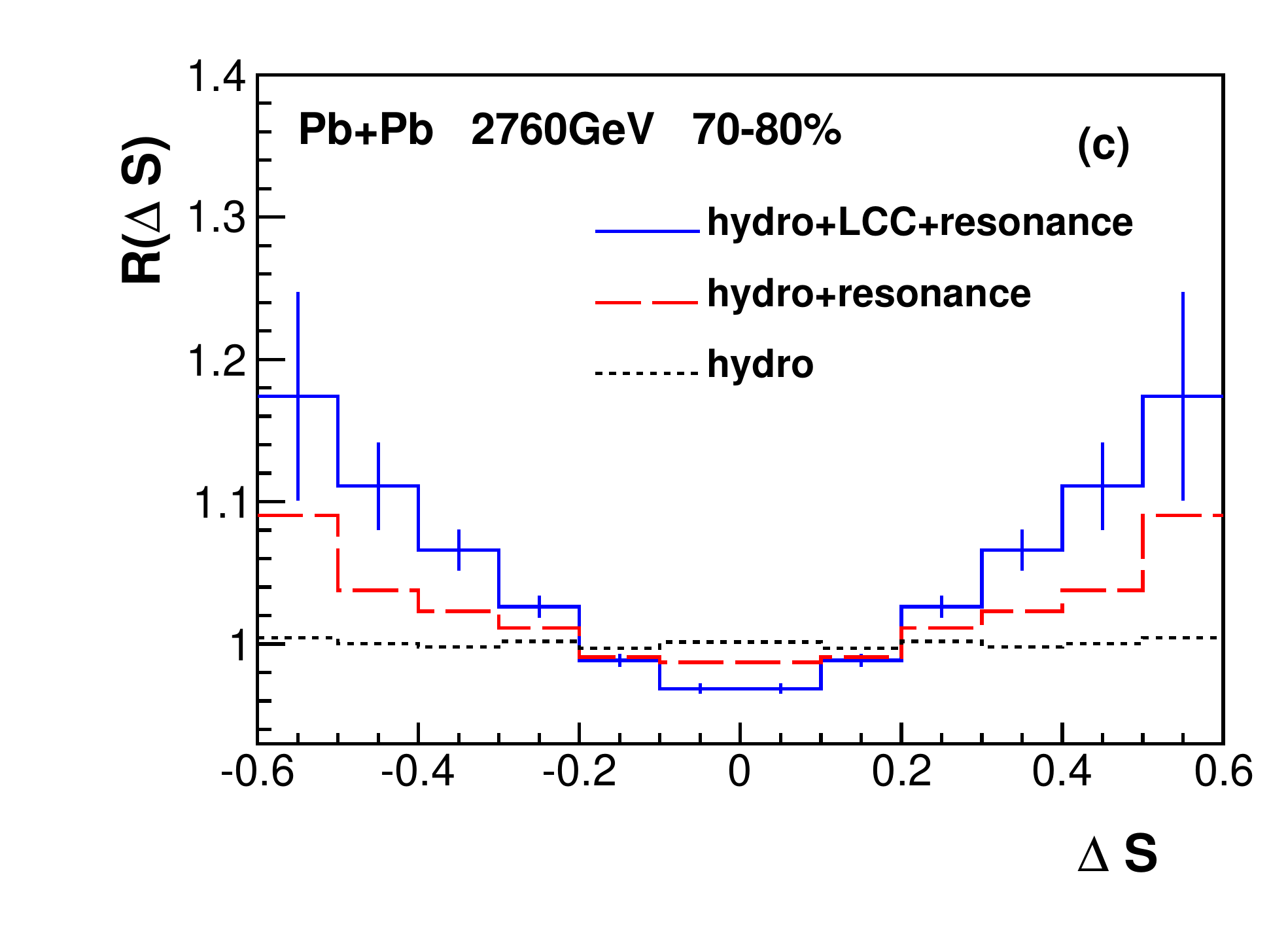}
%\vskip -20mm
\caption{(color online) The correlator $R(\Delta S)$ (Eq.~\ref{eq:Rds})
 for Pb+Pb collisions at $\sqrt{s}=2760$~GeV with $0-5$\% 
(panel (a)), $30-40$\% (panel (b)), and $70-80$\% centrality (panel (c)). 
The dotted lines denote the results for primordial particles only, with 
no charge dependent correlations. The dashed lines represent the results
 of the simulations including particles from resonance decays.
The solid lines represent the results of the model with local charge 
conservation and resonance decays included.  For clarity,
 in all figures statistical errors are shown only for one of the calculations.
\label{fig:pbpb3040}}
\end{figure}

Simulations are performed in a $3+1$-dimensional viscous hydrodynamic model
 \cite{Schenke:2010rr,Bozek:2011ua}. 
After the expansion of the fireball, particles 
are emitted statistically from the freeze-out hypersurface, defined as 
the surface of constant temperature $T=150$~MeV. In the standard 
implementation,
particles are emitted independently \cite{Chojnacki:2011hb}. 
For particles directly emitted from the
 freeze-out hypersurface (primordial particles) no charge
 dependent correlations are present.
  Charge correlations are build in  only from
subsequent decay of resonances. In the model with local charge conservation 
particles are emitted in pairs from the same fluid element, a particle and the 
corresponding antiparticle. The particle and antiparticle share the common 
flow velocity of the fluid and have an independent thermal component of the 
momentum.
This simple model encompasses  the collimation effect between 
opposite  charges 
from the collective flow \cite{Bozek:2012en}.

 The calculation of the correlation sensitive to the chiral magnetic effect requires  a large statistics. In the following I show selected centralities for Pb+Pb collisions 
at $\sqrt{s}=2760$~GeV and p+Pb collisions at $\sqrt{s}=5020$~GeV,
 both using a quark Glauber Monte Carlo model for the hydrodynamic 
initial conditions \cite{Bozek:2017elk}, 
as well as for Au+Au collisions at $\sqrt{s}=200$~GeV 
  using a nucleon Glauber model for initial conditions 
\cite{Bozek:2011ua}.
All the calculations are performed in two versions, one with charge 
 correlations from resonance decays only  and the other including also
 the local charge conservation effect. The details of the initial conditions and of the hydrodynamic modeling are not essential for the present study as long long
as the spectra and the average harmonic flow coefficients are reproduced.
The charge splitting background effects discussed in this paper involve
phenomena happening at the freeze-out and after. The hydrodynamic evolution is needed to obtain a realistic freeze-out hypersurface and flow. 
Alternatively, a simple blast-wave ansatz has been successfully 
used instead \cite{Schlichting:2010qia}.

\begin{figure}[tb]
%\begin{center}
\includegraphics[width=.45 \textwidth]{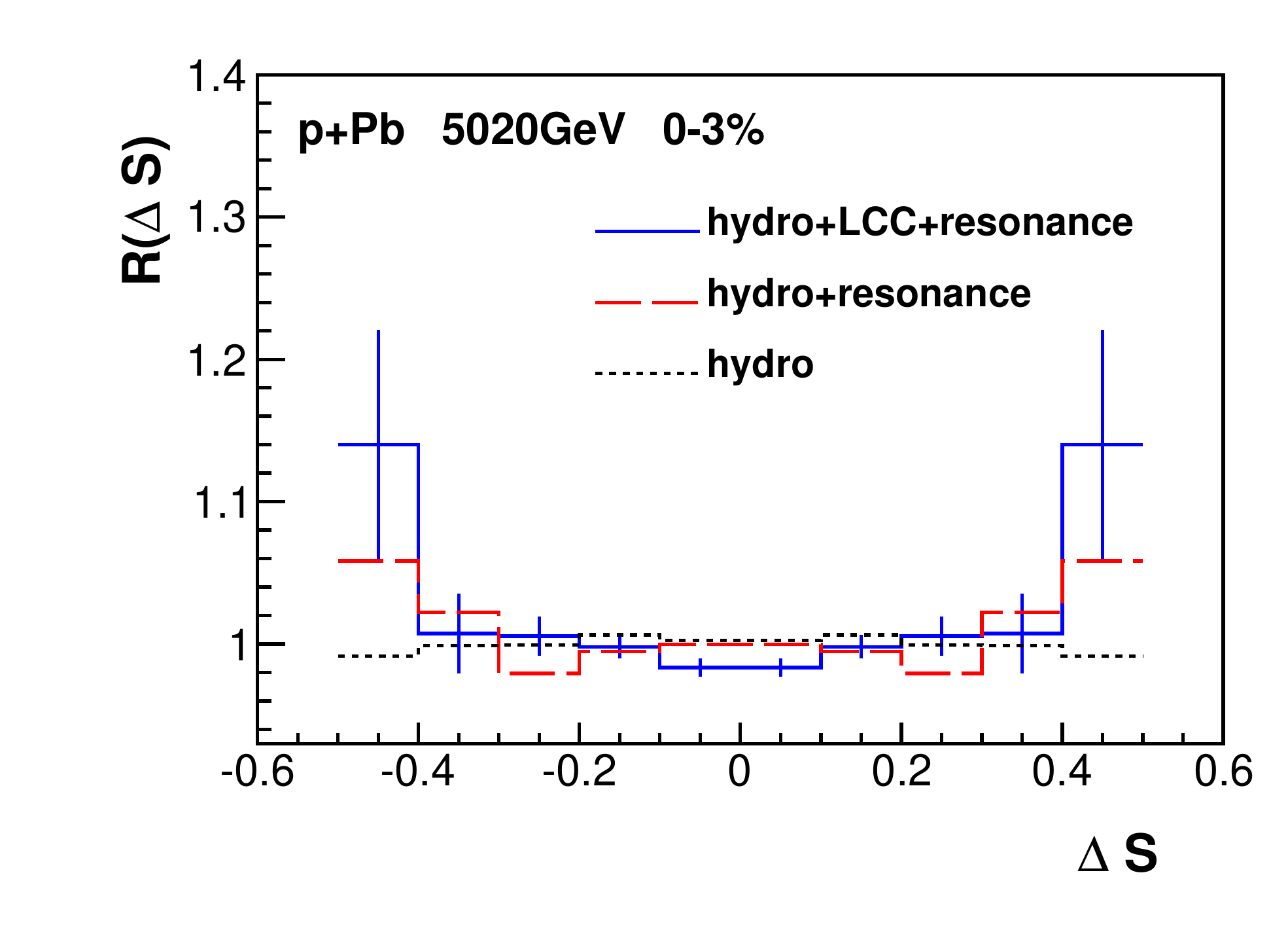}
%%\vspace{-10mm}
\caption{(color online) Same as in Fig. \ref{fig:pbpb3040} but for p+Pb collisions at $\sqrt{s}=5020$~GeV with centrality $0-3$\%.
\label{fig:Rppb}}
\end{figure}

\begin{figure}[tb]
%\begin{center}
\includegraphics[width=.45 \textwidth]{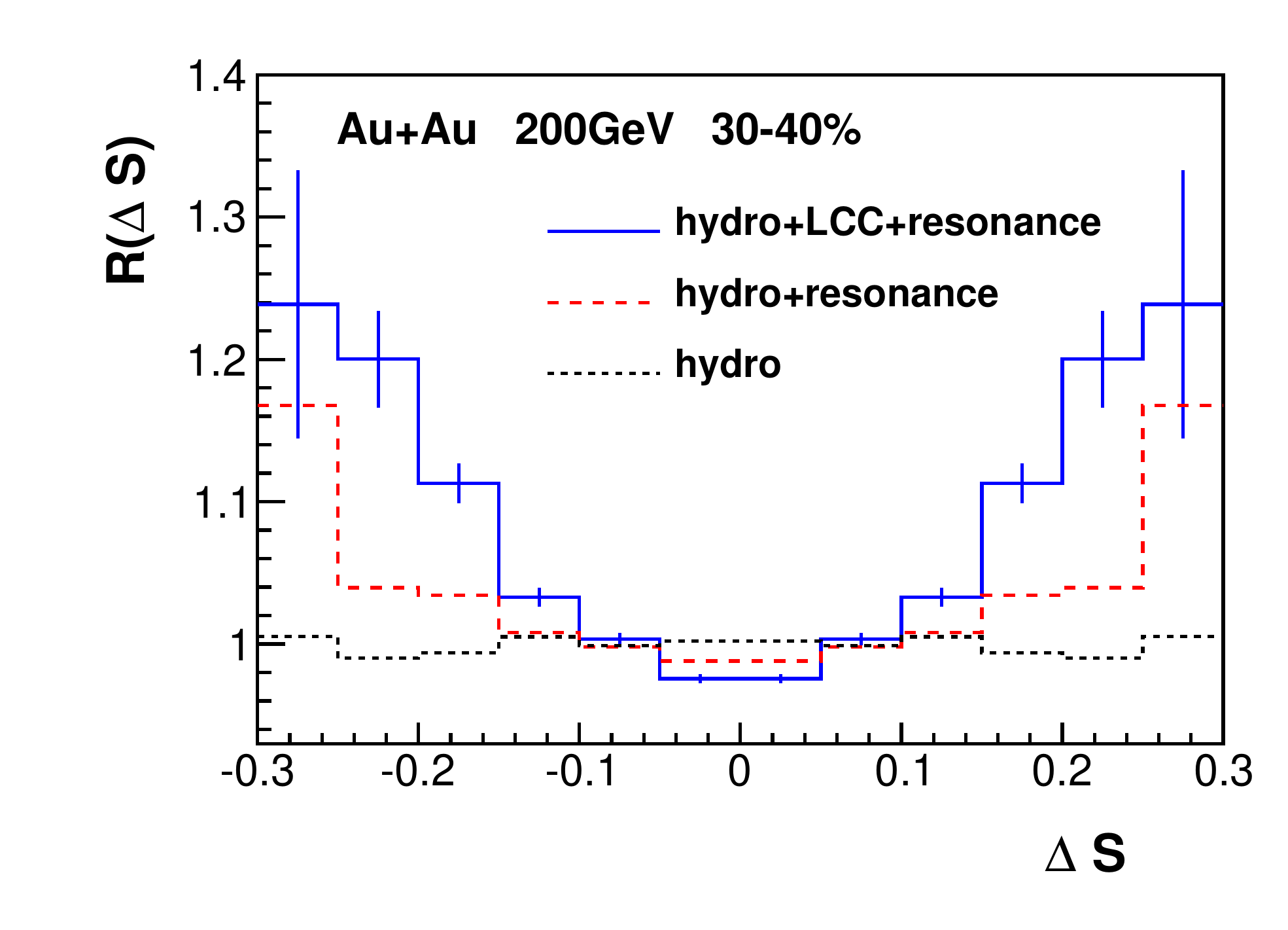}
%\vspace{-10mm}
\caption{(color online) Same as in Fig \ref{fig:pbpb3040} but for Au-Au collisions at $\sqrt{s}=200$~GeV and centrality $30-40$\%.
\label{fig:Rauau}}
\end{figure}

The chiral magnetic effect leads to a charge separation along the direction of 
the magnetic field. As the direction of the magnetic field 
is perpendicular to the 
direction of the second order event plane $\Psi_2$, the presence of topological 
effects should lead to an increase of the magnitude of the charge splitting 
projected on a direction perpendicular to $\Psi_2$ \cite{Ajitanand:2010rc}
\begin{equation}
\Delta S =\frac{ \sum_{i=1}^p \sin(\phi_i -\Psi_2)}{p}-\frac{ \sum_{i=1}^m \sin(\phi_i -\Psi_2)}{m} \ .
\label{eq:DS}
\end{equation}
The sums go over all the $p$ positive and $m$ negative
 charges in the 
acceptance region ($\eta|<1$ , $0.15$~GeV$<p_\perp<2$~GeV). 
The magnitude and the sign of 
$\Delta S$ vary from event to event. The distribution $N(\Delta S)$ 
of $\Delta S$ is constructed. From real events reshuffled events are 
generated by reshuffling the charges of particles in the acceptance region.
The corresponding distribution is $N_{sh}(\Delta S)$.
The ratio of the two distributions is
\begin{equation}
C(\Delta S)=\frac{N(\Delta S)}{N_{sh}(\Delta S)} \ .
\end{equation}
The same ratio is constructed for 
the charge splitting projected on the direction perpendicular to the magnetic
field
\begin{equation}
\Delta S_\perp =\frac{ \sum_{i=1}^p \cos(\phi_i -\Psi_2)}{p}-\frac{ \sum_{i=1}^m \cos(\phi_i -\Psi_2)}{m} \  ,
\end{equation}
with the corresponding correlation $C_\perp(\Delta S_\perp)$.
Finally, the correlator involving the ratio of the two correlations is calculated
\cite{Magdy:2017yje}
\begin{equation}
R(\Delta S)=\frac{C(\Delta S)}{C_\perp(\Delta S)} \ .
\label{eq:Rds}
\end{equation}
The authors of Ref. \cite{Magdy:2017yje}
 notice that  in a model with a chiral magnetic effect and in the preliminary STAR data 
$R(\Delta S)$ has a convex shape, while in other models studied it has a 
concave shape.

The  correlator $R(\Delta S)$ in Pb+Pb collisions is shown in  Fig \ref{fig:pbpb3040}. In all panels three results are compared, one
 using a model with local charge conservation and resonance decays, one from 
a model with resonances only, and one from a model
 where only primordial particles are taken. In the calculation the second order event-plane direction $\Psi_2$ is reconstructed from combined events, involving many statistical events generated from the same freeze-out hypersurface. Thus the event plan resolution is close to one. However, 
the charge splitting $\Delta S$ (Eq. \ref{eq:DS}) is calculated from
 real events with a realistic multiplicity.

As expected, primordial particles
 show no charge dependent correlations. The other two calculations show a
 convex shape
for the function $R(\Delta S)$. The model with local charge conservation shows a stronger charge dependence of the correlator than the model with resonances only, except for the centrality $0-5$\% where the two are compatible within
 errors. The convex-like deviation of the  correlator $R(\Delta S)$ from $1$ is the strongest in semi-central and peripheral collisions. Due to the elliptic flow, the azimuthal 
dependence of the fluid flow velocity is the strongest in these cases. 
The stronger the  flow the more collimated are opposite-charged particle pairs
from resonance decays and from the local charge conservation. 
Qualitatively, similar results are obtained for  p+Pb   (Fig. \ref{fig:Rppb})
 and  Au+Au collisions (Fig. \ref{fig:Rauau}).
 The range of $\Delta  S$ increases with decreasing average multiplicity  in the events.

\section{Distribution of the charge imbalance in- and out-of-plane}

\begin{figure}[tb]
%\begin{center}
\includegraphics[width=.45 \textwidth]{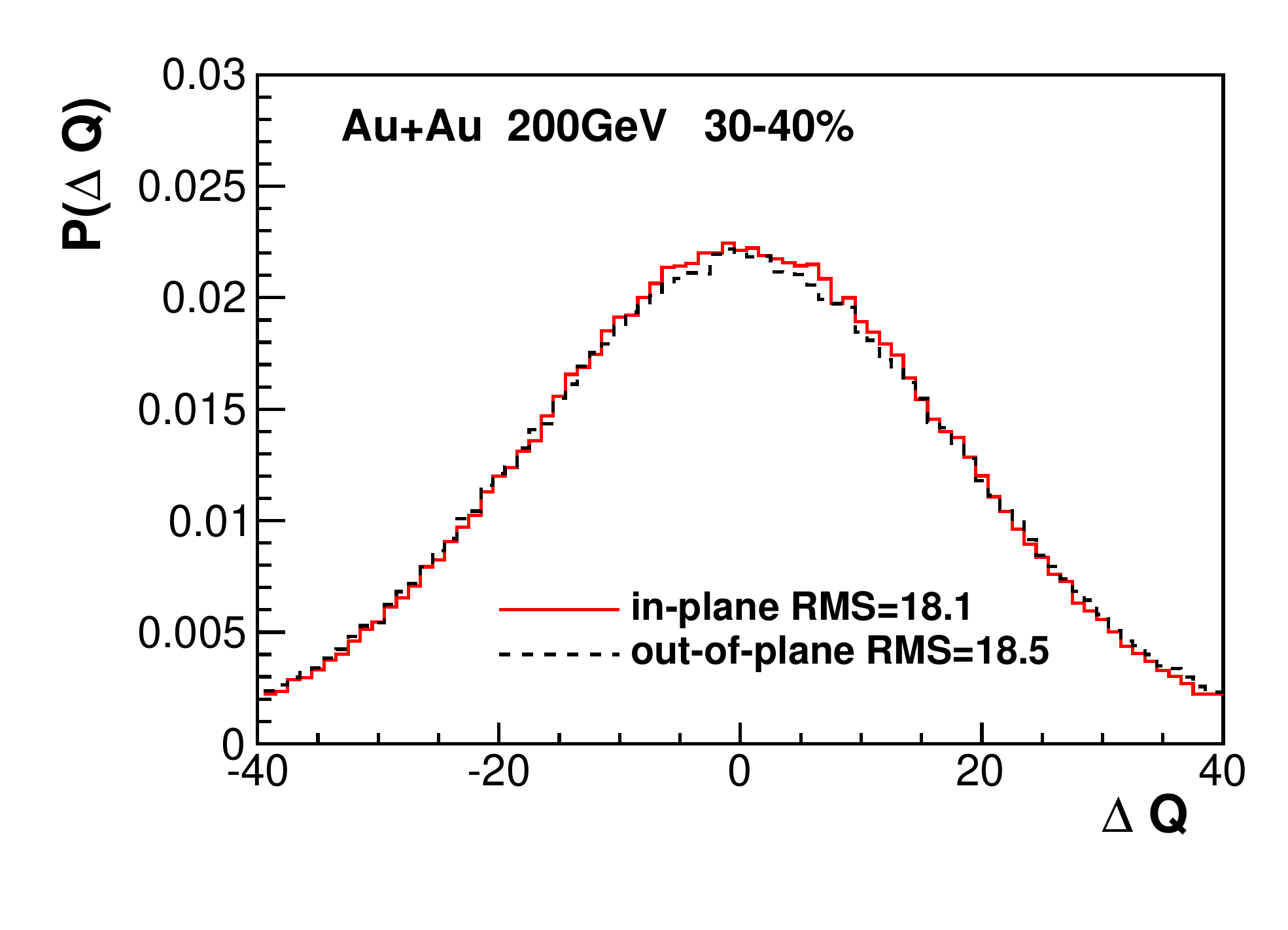}
%\vspace{-10mm}
\caption{(color online) In- and out-of-plane charge imbalance distributions
 for Au+Au collisions at $\sqrt{s}=200$~GeV and centrality $30-40$\%.
\label{fig:qdist}}
\end{figure}

The STAR Collaboration presented a related study, comparing the distribution of the charge imbalance in the  directions in- and out-of-plane
\cite{Adamczyk:2013hsi}. The in-plane charge imbalance is defined as
\begin{equation}
\Delta Q_{in} = Q_I -Q_{III} \ ,
\end{equation}
where $Q_I$ and $Q_{III}$ denote the total charge of particles registered in the 
quadrants $\Psi_2-\frac{\pi}{4}<\phi<\Psi_2+\frac{\pi}{4}$ and
  $\Psi_2+\frac{3\pi}{4}<\phi<\Psi_2+\frac{5\pi}{4}$.
Analogously for the out-of-plane direction
\begin{equation}
\Delta Q_{out} = Q_{II} -Q_{IV} \ ,
\end{equation}
where $Q_{II}$ and $Q_{IV}$ denote the total charge of particles registered
 in the 
quadrants $\Psi_2+\frac{\pi}{4}<\phi<\Psi_2+\frac{3\pi}{4}$ and
  $\Psi_2+\frac{5\pi}{4}<\phi<\Psi_2+\frac{7\pi}{4}$.

The event-by-event distributions of the charge imbalances
$\Delta Q_{in}$ and $\Delta Q_{out}$ are shown in Fig. \ref{fig:qdist}. 
The in-plane charge imbalance is narrower than the out-of-plane one. The 
relative difference of the rms values for the two distributions
\begin{equation}
\frac{ RMS_{out}-RMS_{in}}{(RMS_{out}+RMS_{in})/2)}=0.022
\end{equation}
is close to the experimental value 0.019 \cite{Adamczyk:2013hsi}.

\section{Charge splitting distribution with 
respect to the third order event plane}

\begin{figure}[tb]
%\begin{center}
%\vskip -60mm
\includegraphics[width=.45 \textwidth]{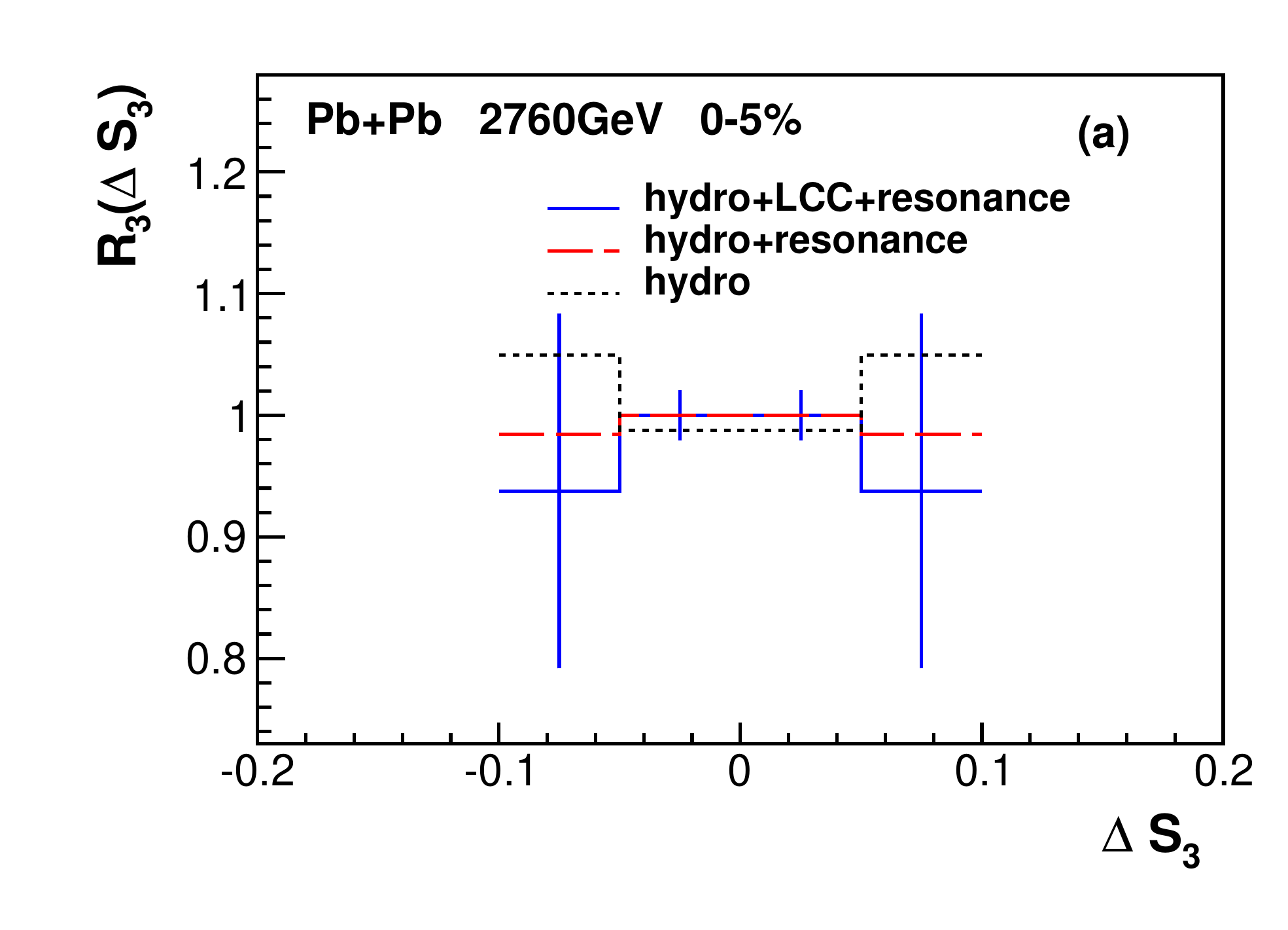}

\includegraphics[width=.45 \textwidth]{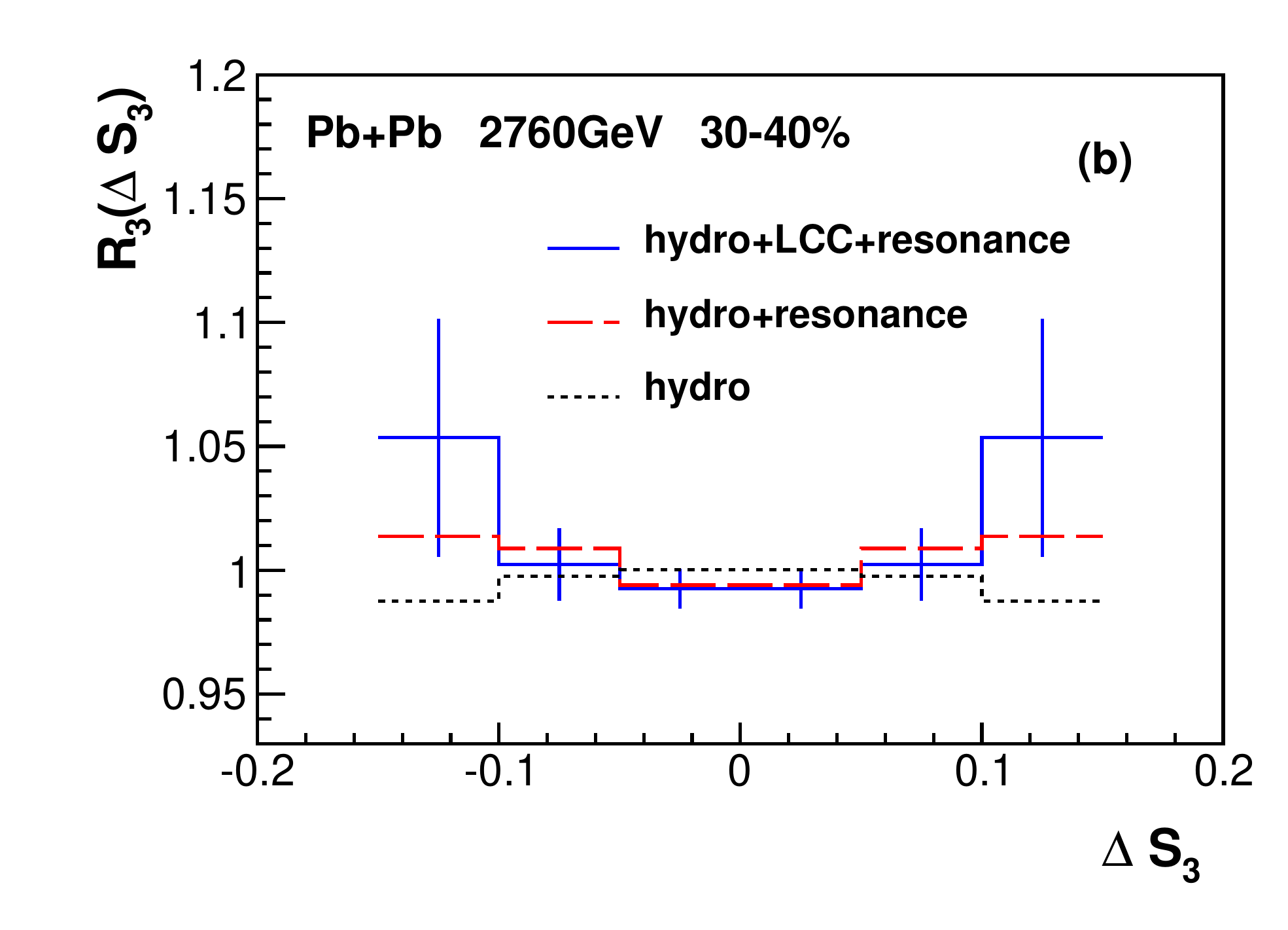}

\includegraphics[width=.45 \textwidth]{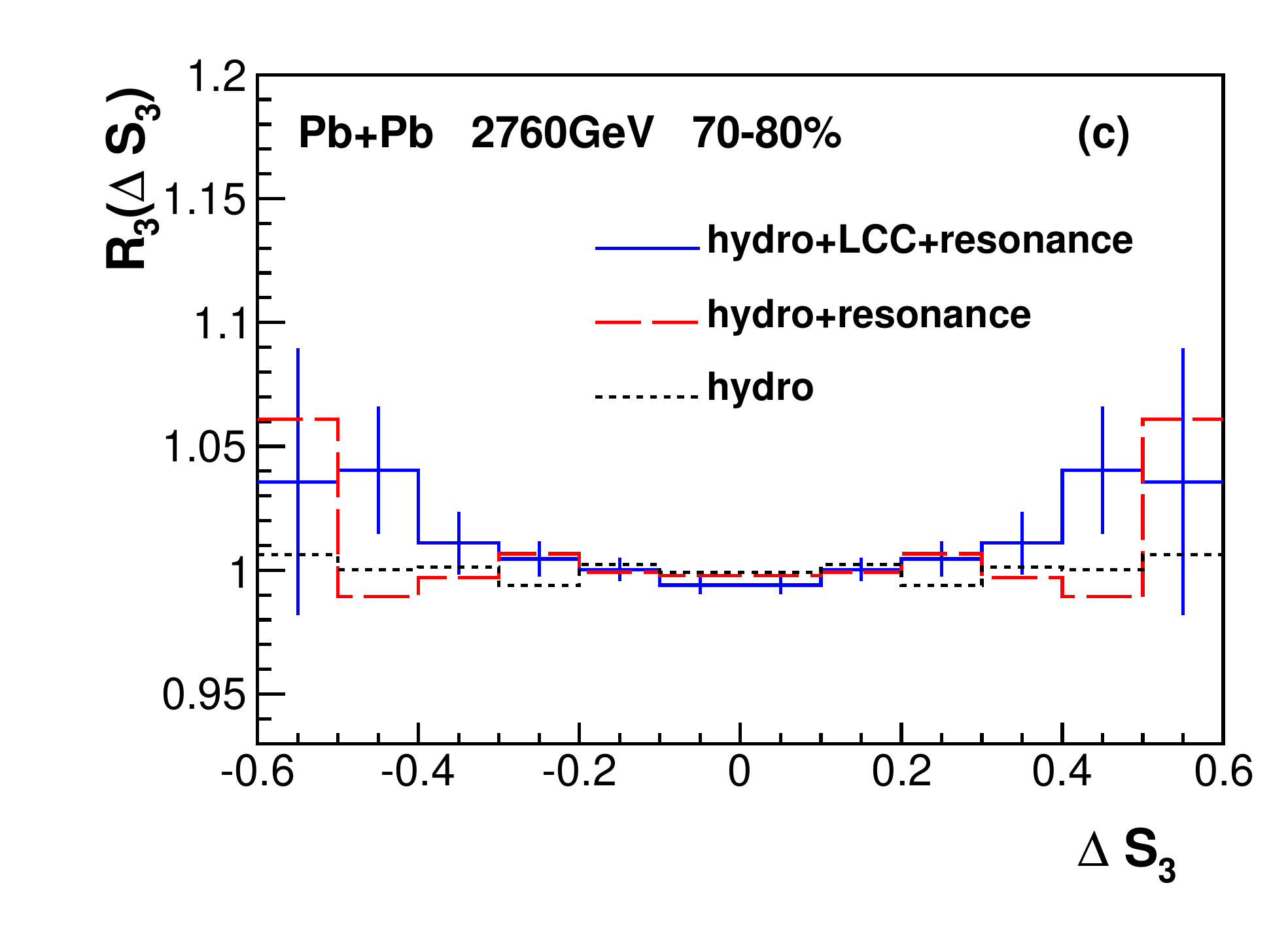}
%\vskip -20mm
\caption{(color online) 
\label{fig:R3LHC} Same as in Fig. \ref{fig:pbpb3040} but for the third order 
event plane (Eq. \ref{eq:Rds3}). }
\label{fig:pbpb3}
\end{figure}

\begin{figure}[tb]
%\begin{center}
\includegraphics[width=.45 \textwidth]{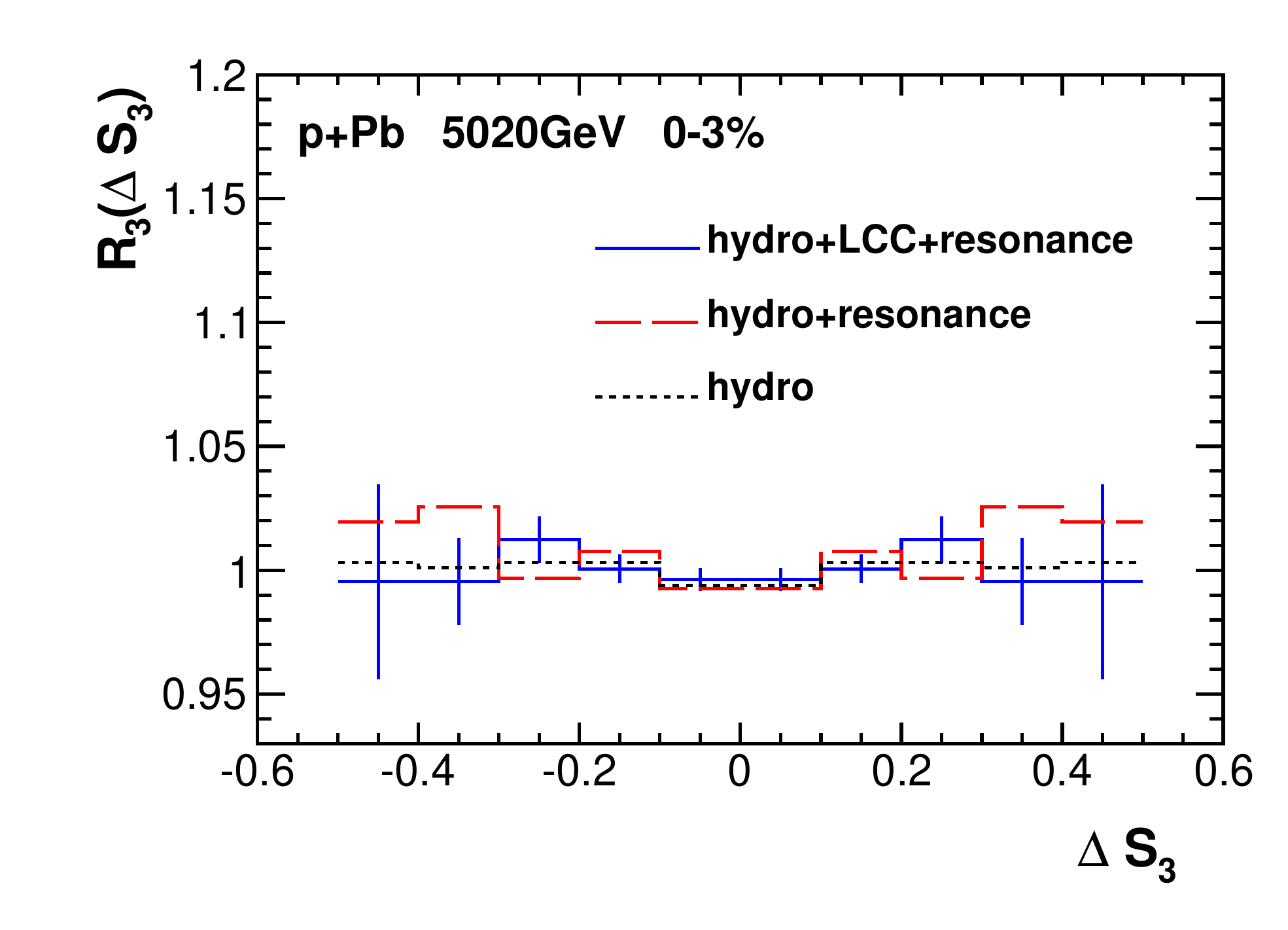}
%\vspace{-10mm}
\caption{(color online) Same as in Fig. \ref{fig:pbpb3} but for p+Pb collisions at $\sqrt{s}=5020$~GeV with centrality $0-3$\%. 
\label{fig:R3ppb}}
\end{figure}

\begin{figure}[tb]
%\begin{center}
\includegraphics[width=.45 \textwidth]{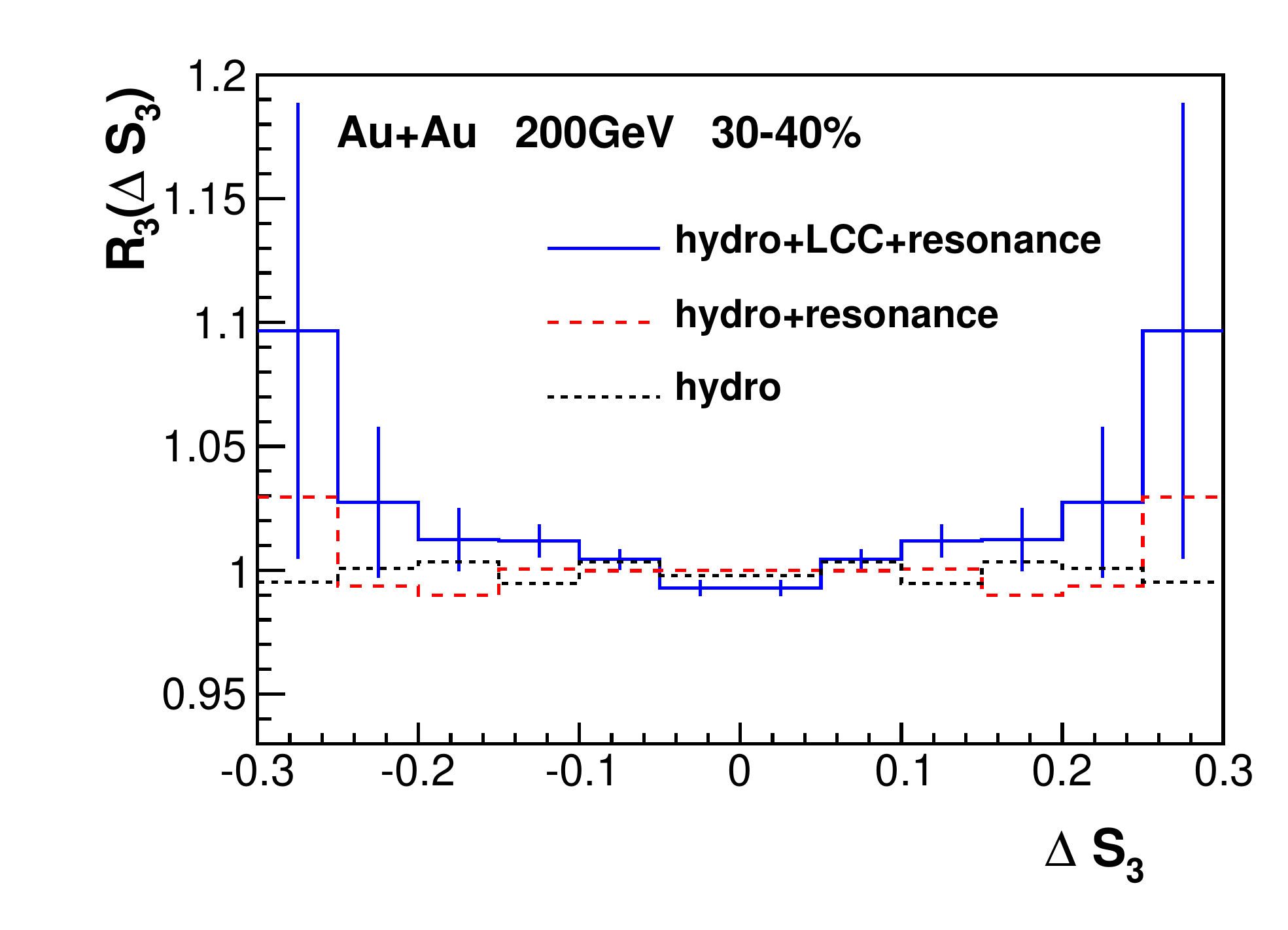} 
%\vspace{-10mm}
\caption{(color online)  Same as in Fig \ref{fig:pbpb3} but for Au+Au collisions at $\sqrt{s}=200$~GeV and centrality $30-40$\%.
\label{fig:R3auau}}
\end{figure}

The azimuthal asymmetry of the collective flow contains a third order
 component, the triangular flow. The charge dependent correlations 
depend on the flow and should exhibit a modulation with respect to the
angle of the  third order flow component $\Psi_3$. The  directions 
of the minimal flow 
are located at $\Psi_3\pm \frac{\pi}{3}$ and   $\Psi_3+\pi$.
The projection of the charge splitting on the direction  of minimal flow 
can be defined as
\begin{equation}
\Delta S_3 =\frac{ \sum_{i=1}^p \sin\left(\frac{3}{2}(\phi_i -\Psi_3)\right)}{p}-\frac{ \sum_{i=1}^m \sin\left(\frac{3}{2}(\phi_i -\Psi_3)\right)}{m}
\label{eq:DS3}
\end{equation}
and analogously for the directions along the flow
\begin{equation}
\Delta S_{\perp,3} =\frac{ \sum_{i=1}^p \cos\left(\frac{3}{2}(\phi_i -\Psi_3)\right)}{p}-\frac{ \sum_{i=1}^m \cos\left(\frac{3}{2}(\phi_i -\Psi_3)\right)}{m} \ .
\end{equation}
The events-by-event distributions of $\Delta S_3$ and $\Delta S_{\perp,3}$ are
constructed for the real and reshuffled events to obtain the distributions $C_3(\Delta S_3)$ and $C_{\perp,3}(\Delta S_{\perp,3})$.
Finally the correlator for the charge splitting with respect to the third 
order flow is calculated
\begin{equation}
R_3(\Delta S_3)=\frac{C_3(\Delta S_3)}{C_{\perp,3}(\Delta S_3)} \ .
\label{eq:Rds3}
\end{equation}

The results for the third order correlator are shown in Figs. \ref{fig:pbpb3}, 
\ref{fig:R3ppb}, and \ref{fig:R3auau}. The convex deviation of the third order correlator $R_3(\Delta S_3)$  from $1$ is
visible for semi-central and peripheral collisions of heavy ions, although the
effect is less pronounced than for the second order correlator $R(\Delta S)$.
For central Pb+Pb collisions and for p+Pb collisions no significant deviation of the correlator from $1$ can be evidenced.

\section{Event plane resolution effects}

\begin{figure}[tb]
%\begin{center}
%\vskip -60mm
\includegraphics[width=.45 \textwidth]{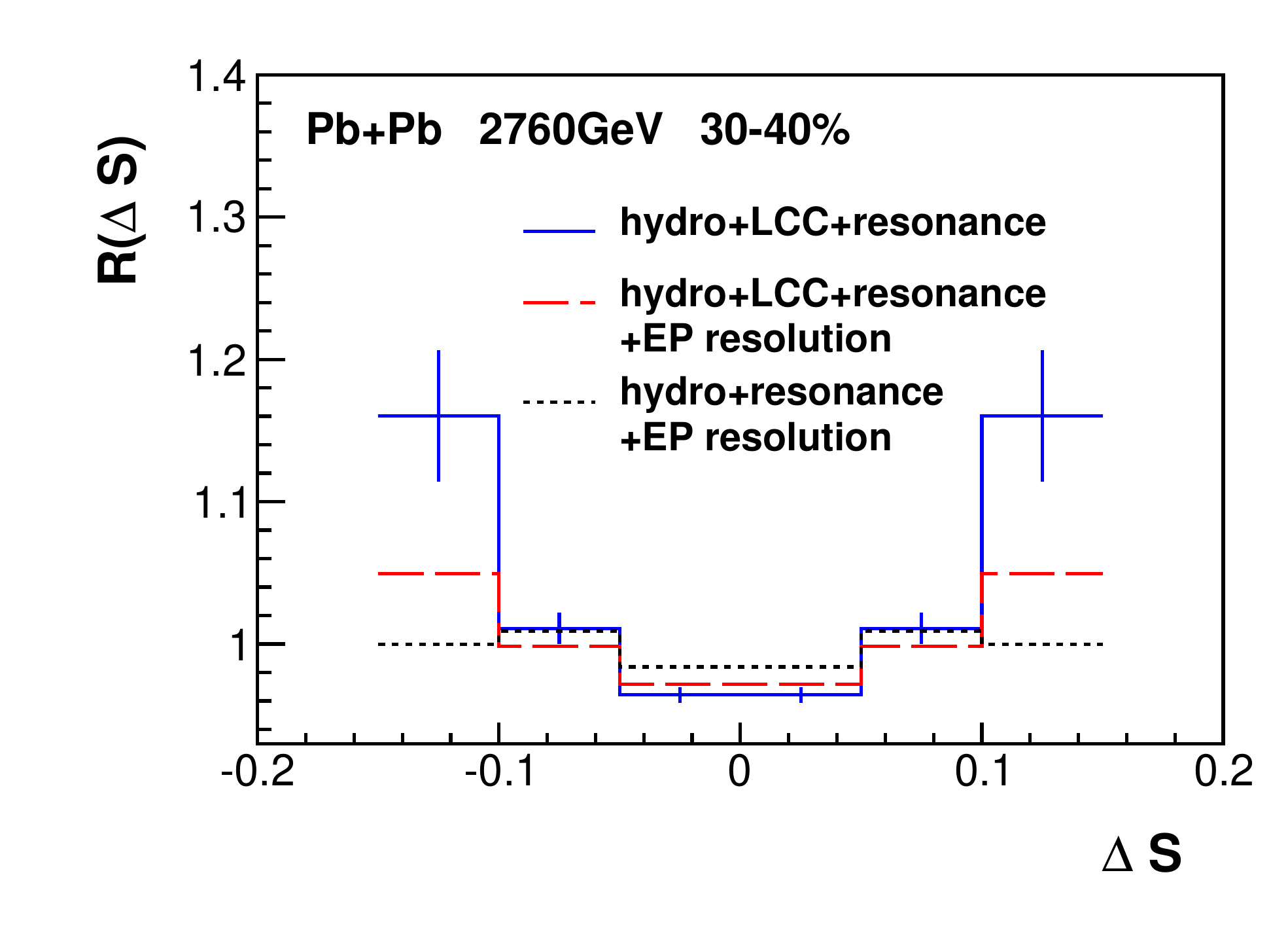}
%\vskip -20mm
\caption{(color online) The correlator $R(\Delta S)$ in Pb+Pb collisions
 at $\sqrt{s}=2070$~Gev and centrality $30-40$\%. 
The solid line denotes the result with a perfect event-plane
 resolution, and the dashed and dotted lines for the event plane defined 
from charged particles with $2<|\eta|<4$. The results from the model including resonance decays is denoted with the dotted line and the calculation with resonances and local charge conservation with the  dashed line.
\label{fig:realpbpb}}
\end{figure}

\begin{figure}[tb]
%\begin{center}
\includegraphics[width=.45 \textwidth]{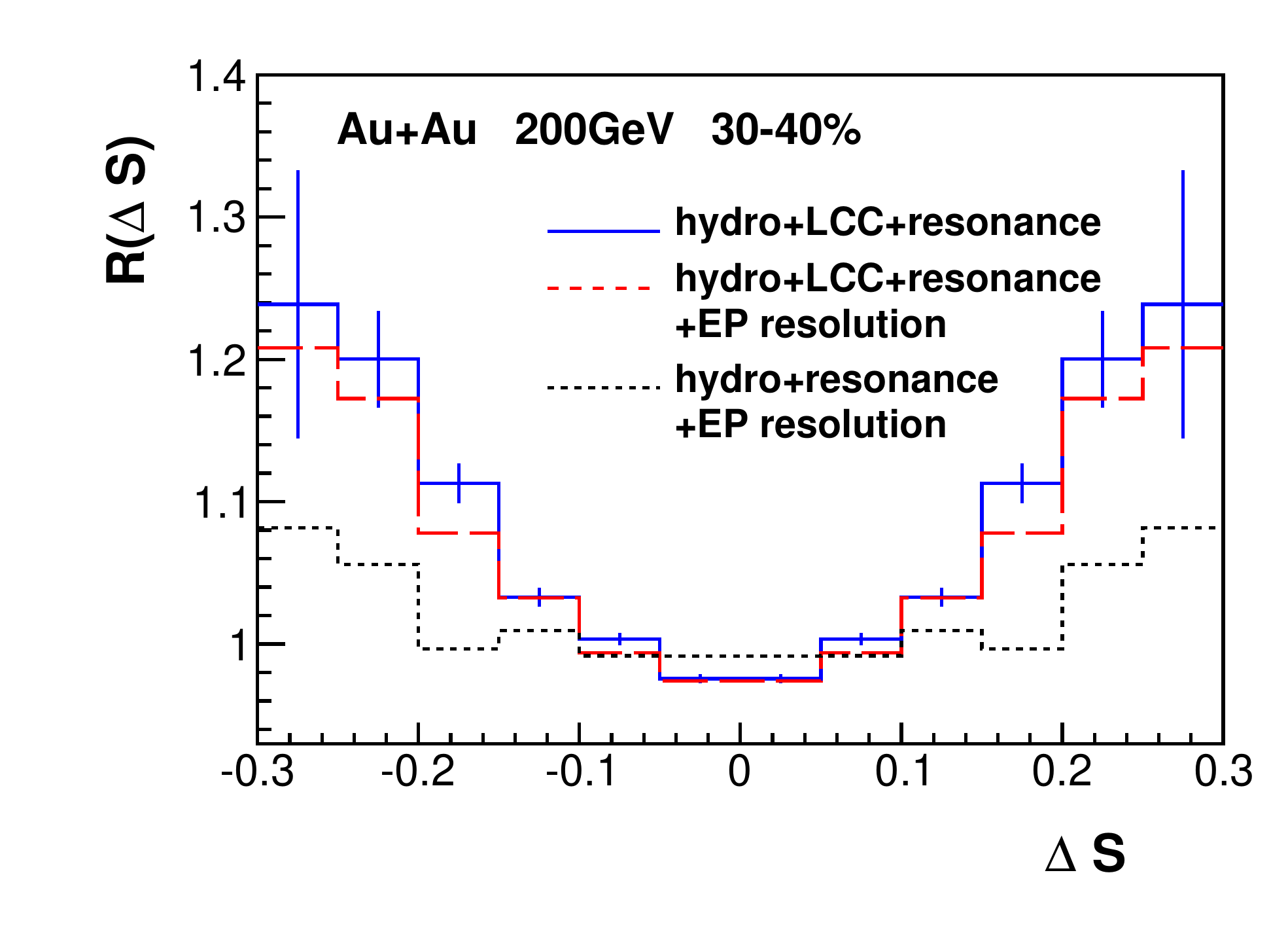}
%\vspace{-10mm}
\caption{(color online)  Same as in Fig. \ref{fig:realpbpb} but for Au-Au collisions at $\sqrt{s}=200$~GeV and centrality $30-40$\%.
\label{fig:Rrealauau}}
\end{figure}

\begin{figure}[tb]
%\begin{center}
%\vskip -60mm
\includegraphics[width=.45 \textwidth]{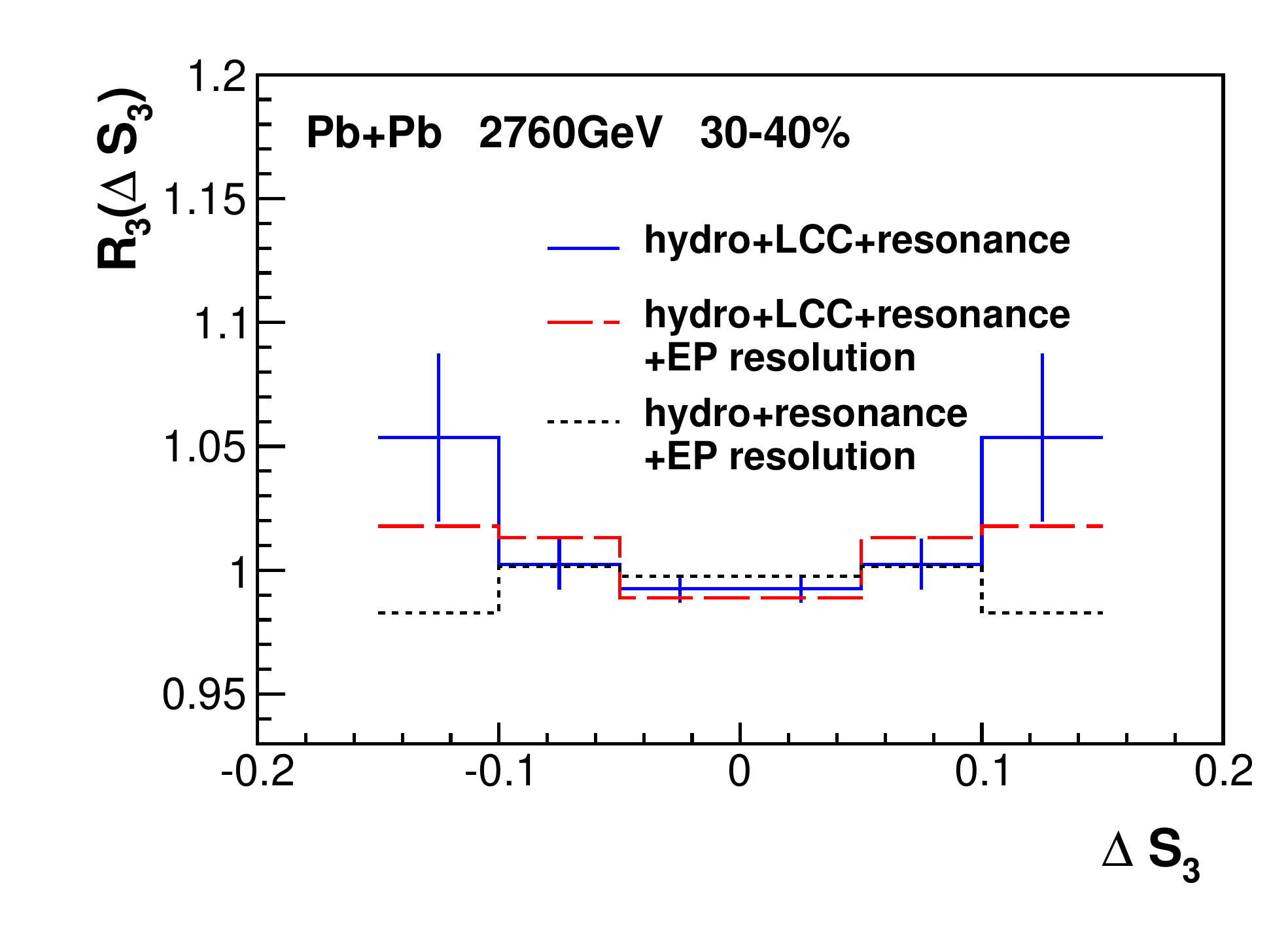}
%\vskip -20mm
\caption{(color online)  Same as in Fig. \ref{fig:realpbpb} but for the correlator $R_3(\Delta S)$ defined with respect to the third order event plane.
\label{fig:Rreal3LHC}}
\end{figure}

\begin{figure}[tb]
%\begin{center}
\includegraphics[width=.45 \textwidth]{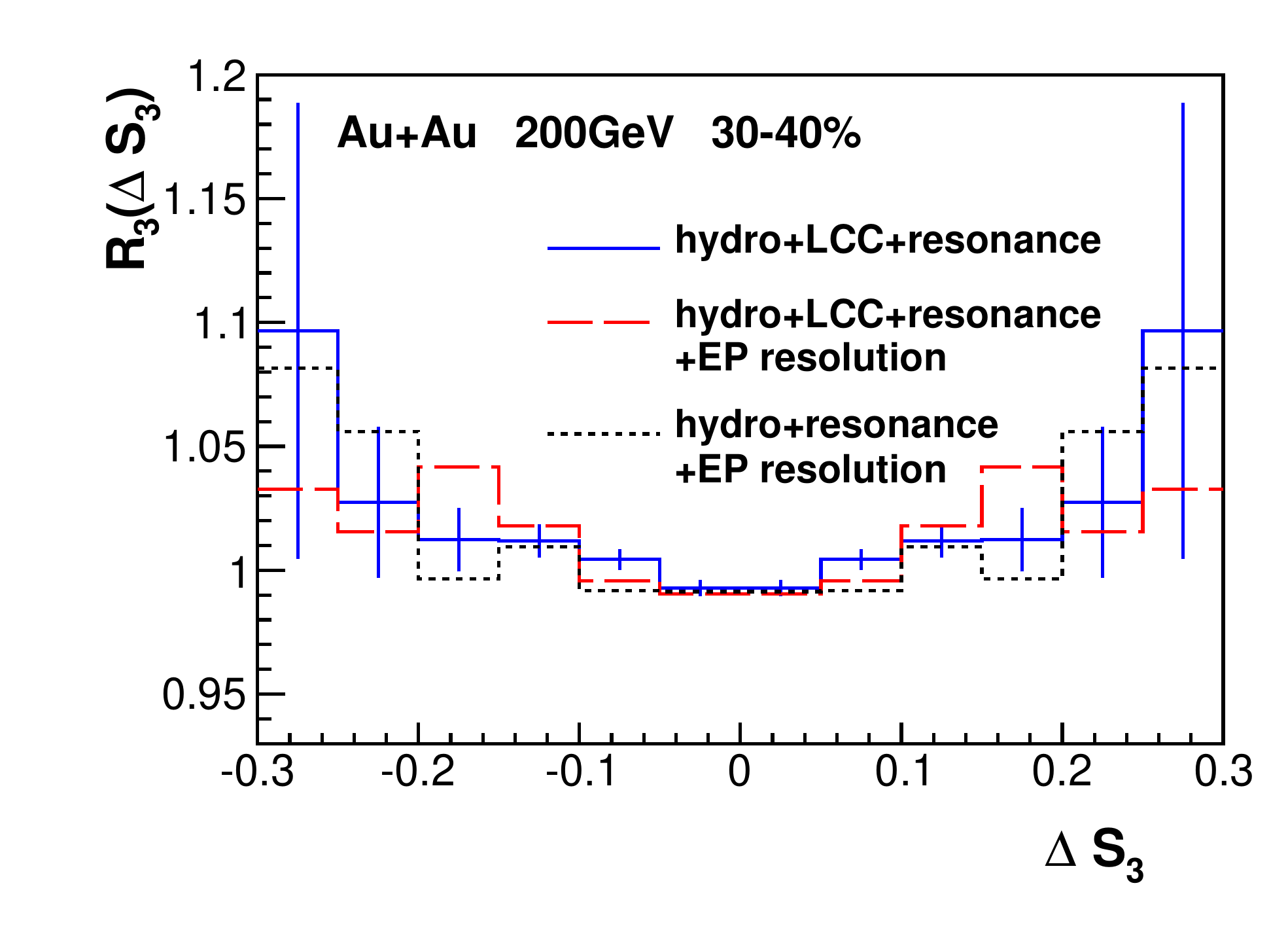}
%\vspace{-10mm}
\caption{(color online)  Same as in Fig. \ref{fig:Rrealauau} but for the correlator $R_3(\Delta S)$ defined with respect to the third order event plane.
\label{fig:Rreal3auau}}
\end{figure}

The calculation of the charge splitting with respect to $\Psi_2$ 
and $\Psi_3$ requires the reconstruction of the corresponding event planes. 
In the model the deviation of the reconstructed events plane $\Psi_{2,3}$ 
from the true flow direction $\Psi_{2,3}^{flow}$ can be estimated.
The actual result depends on the particular definition of the event plane in the
experimental procedure.  For illustration, one case is studied,
 defining the event plane from $50$\% of charged particles
 with $2<|\eta|<4$ and $0.15$~GeV$<p_\perp<2$~GeV. Finite event-plane resolution
is expected to reduce to relative differences between correlations 
for the  in- and out-of-plane directions.

Another aspect of the procedure, that influences the $\Delta S$ 
and $\Delta S_3$ distributions is the choice of the acceptance window for charged
 particles used in the analysis   and the corresponding efficiency.
 By reducing the acceptance window or by lowering the efficiency the 
distributions 
of $\Delta S$ and $\Delta S_3$ get broader. In this section we keep 
the same acceptance window as before ($|\eta|<1, \ 0.15$~GeV$<p_\perp<2$~GeV), 
but with an efficiency of $80$\%.

The results of the calculation for $R(\Delta S)$ including the effect of finite
event-plane resolution are plotted in Figs. 
\ref{fig:realpbpb} and \ref{fig:Rrealauau}. 
Qualitatively, the results are not modified by a finite event-plane resolution,
but the deviation of the correlator from $1$ is smaller. 
The third order correlator $R_3(\Delta S_3)$ is 
more sensitive to a finite event-plane resolution.  The signal is weaker
 for $R_3$ and the event-plane resolution is usually much poorer
 for the triangular
 than for the elliptic flow. The deviation of the third order correlator
from $1$ is strongly reduced, or could be even washed out by event-plane resolution effects (Figs. \ref{fig:Rreal3LHC} and \ref{fig:Rreal3auau}). 
In order to  measure 
$R_3(\Delta S_3)$ a  setup allowing for a good event-plane resolution 
should be used.

\section{Conclusions}

A correlator $R(\Delta S)$  (Eq.~\ref{eq:Rds}) comparing 
the event-by-event charge splitting in the directions along 
and perpendicular to
the magnetic field in heavy-ion collisions has been proposed as 
a sensitive probe of the chiral magnetic effect \cite{Magdy:2017yje}.
The presence of topological domains in the deconfined phase should lead
to an enhanced charge splitting in the direction of the magnetic field giving
a convex shape of $R(\Delta S)$. 
I show that qualitatively the same behavior can be reproduced due to standard
charge correlations, such as resonance decays and local charge conservation.
The results of 3+1-dimensional hydrodynamic simulations in Pb+Pb, p+Pb, and 
Au-Au collisions show all a convex shape of $R(\Delta S)$.

The results show that a convex shape of the
charge splitting observable $R(\Delta S)$ cannot be used and unambiguous
 evidence
of the chiral magnetic effect in heavy-ion collisions.
 The background effects from standard phenomena
give a very large contribution to this observable.
 It would be very challenging to calculate reliably and subtract the
background contribution from the measured correlator in order to extract a
 possible signal of the chiral magnetic effect.

If charge dependent  correlations are due to resonance decays and/or
 local charge conservation  the charge splitting should have a third order
azimuthal dependence from the triangular flow. Simulations predict such 
an effect.
This new observable could be 
measured in experiments that have a good   event-plane
 resolution. It would give an additional constraint on the background effects
present in the observables sensitive to the chiral magnetic effect.

\begin{acknowledgments}

The author thanks Adam Bzdak, Sandeeep Chatterjee, and Roy Lacey for discussions.
Research supported by the Polish Ministry of Science and Higher Education (MNiSW), by the National
Science Centre grant  2015/17/B/ST2/00101, as well as by PL-Grid Infrastructure. 

\end{acknowledgments}

\bibliography{../hydr}

\end{document}